\documentclass[aip,reprint,apl]{revtex4-1}
\usepackage{graphicx}
\usepackage{amsmath}

\usepackage{textcomp}
\newcommand{\murm}{%
	\mathchoice
	{\hbox{\normalsize\textmu}}
	{\hbox{\normalsize\textmu}}
	{\hbox{\scriptsize\textmu}}
	{\hbox{\tiny\textmu}}%
}

\begin{document}
	
	\title{Spin wave propagation in corrugated waveguides} 
	
	
	
	\author{Igor Tur\v{c}an}
	\email{igor.turcan@vutbr.cz}
	\author{Luk\'{a}\v{s} Flaj\v{s}man}%
	\author{Ond\v{r}ej Wojewoda}
	\affiliation{ 
		CEITEC BUT, Brno University of Technology, 612 00 Brno, Czech Republic
	}%
	\author{V\'{a}clav Rou\v{c}ka}
	\affiliation{Institute of Physical Engineering, Brno University of Technology, 616 69 Brno, Czech Republic}
	\author{Ond\v{r}ej Man}
	\affiliation{ 
		CEITEC BUT, Brno University of Technology, 612 00 Brno, Czech Republic
	}%
	\author{Michal Urb\'{a}nek}
	\email{michal.urbanek@ceitec.vutbr.cz}
	\affiliation{ 
		CEITEC BUT, Brno University of Technology, 612 00 Brno, Czech Republic
	}%
	\affiliation{Institute of Physical Engineering, Brno University of Technology, 616 69 Brno, Czech Republic}

	\date{\today}
	
	\begin{abstract}
		Curvature-induced effects allow us to tailor the static and dynamic response of a magnetic system with a high degree of freedom.  We study corrugated magnonic waveguides deposited on a sinusoidally modulated substrate prepared by focused electron beam deposition. The surface curvature in films with thicknesses comparable to the amplitude of modulation locally modifies the contributions of dipolar and exchange energies and results in an effective anisotropy term which can be tuned on-demand based on the exact geometry. We show, by Brillouin light scattering microscopy, that without the presence of an external magnetic field, spin waves propagate over a distance 10×larger in the corrugated waveguide than in the planar waveguide.  Further, we analyze the influence of the modulation amplitude on spin-wave propagation and conclude that for moderate modulation amplitudes, the spin-wave decay length is not affected. For larger amplitudes, the decay length decreases linearly with increasing modulation. 

	\end{abstract}

	
	\maketitle 
	
	Magnonic circuits are increasingly attracting interest for their possible applications in low-power wave-based computing technology and information processing \cite{Chumak2017,Kruglyak2010}. They are considered to be a potential candidate for beyond-CMOS generation of logic circuits \cite{Chumak2015,Sander2017,Csaba2017,Wang2020} as no Joule heat is produced during operation and a non-linear regime is easily accessible.
	
	One of the elementary premises of complex 2D magnonic circuits is a need for operation in the absence of an external magnetic field. If an external magnetic field is used to stabilize magnetization, even a basic circuit element as spin-wave turn exhibits large dispersion mismatch for regions before and after the turn. Local control of the effective field would allow stabilizing the magnetization of different parts of the magnonic circuit in the desired direction, thus preventing the dispersion mismatch. 
	
	The curvature induced effects \cite{Streubel2016} can be used to tailor static and dynamic response of a magnetic system with a high degree of freedom. One of the possible approaches is to use the surface curvature to control the magnitude and the direction of the magnetic anisotropy \cite{Tretiakov2017}. The forming of the curvature-induced uniaxial magnetic anisotropy has been previously demonstrated in several experiments using ion beam induced erosion \cite{Feng2012,Ki2015,Chen2012,Korner2013}. However, these approaches rely on the irradiation of the whole sample area by a broad ion beam. Similarly, as in the case of the magnetocrystalline anisotropy control via field-sputtering or field-annealing \cite{Chikazumi1956,Takahashi1962}, it is not easy to control the direction of this types of induced anisotropy on small scales. Only recently, there were reported alternative approaches allowing local anisotropy control in thin films via focused ion beam irradiation by exploiting either structural phase transformation \cite{Urbanek2018,Flajsman2020} or induced chemical disorder \cite{Nord2019}. However, these approaches rely on unconventional material systems and thus are not universally applicable.
	
	Recent advances in focused electron beam induced deposition (FEBID) allow nanofabrication of 3D structures with shapes unobtainable by standard lithography approaches \cite{Utke2008,Fernandez-Pacheco2017,Fernandez-Pacheco2020,Huth2012,Al2018}. The move towards the third dimension is already happening also in magnonics \cite{Gubbiotti2019,Sakharov2020}, as it has a great potential with regard to new functionalities.
	
	Here, we study corrugated magnonic waveguides deposited on a sinusoidally modulated substrate prepared by FEBID. The surface curvature in films with thicknesses comparable to the amplitude of modulation locally modifies the contributions of dipolar and exchange energies and results in an effective anisotropy term which can be tuned on-demand based on the exact geometry \cite{Tretiakov2017}. We experimentally reveal how the amplitude of modulation influences the strength of the uniaxial anisotropy and how it affects the spin-wave propagation. Our findings show that the concept of corrugated magnonic waveguides is useful, universal and that the corrugated waveguides can be used as building blocks for spin-wave circuits operating without the need for the external magnetic field.
	
\begin{figure}[hbtp!]
		\centering
		\includegraphics{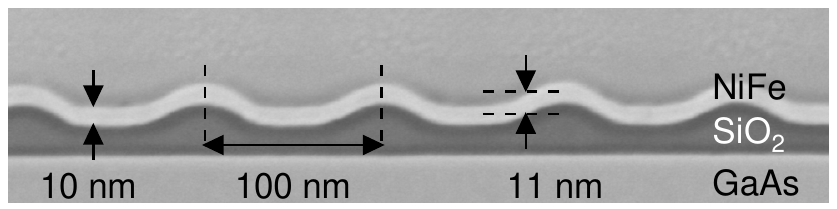}
		\caption{STEM image of the structure prepared with 6,000 scans of the e-beam. The permalloy layer with the nominal thickness of $t=10$\,nm deposited on top of the modulated SiO$_2$ structure is clearly visible. The period of modulation of the corrugated structures is $d=100$\,nm, and the amplitude of modulation (peak-to-peak) for this structure is $A=11$\,nm.}
		\label{Figure1}
	\end{figure}

	Focused electron beam induced deposition, electron beam lithography, e-beam evaporation, and lift-off techniques were used to fabricate samples with corrugated permalloy (Py, Ni$_{81}$Fe$_{19}$) disks and waveguides. All structures were fabricated on GaAs (100) substrate. The sinusoidal surface patterns were prepared in a scanning electron microscope (Tescan LYRA3) equipped with a gas injection system (GIS). We scanned the electron beam with a spot size of 5\,nm in a series of single lines separated by a distance of 100\,nm while introducing PMCPS (2,4,6,8,10-Pentamethylcyclopentasiloxane) precursor into the microscope vacuum chamber. The electrons decomposed the gas molecules of the precursor, and the silicon dioxide structure with the desired sinusoidal shape grew directly on the substrate surface. The acceleration voltage was set to 30\,kV and the beam current to 542\,pA. The base pressure in the chamber before introducing the PMCPS precursor was $1.8\times10^{-4}$\,Pa. We varied the number of scans from 4,000 to 10,000, which resulted in the modulation amplitudes ranging from 4\,nm to 19\,nm. Then, in the next steps, we used e-beam lithography with the PMMA resist to define the required shapes of corrugated magnetic structures [disks with diameter $D=7\,\murm$m and $(2\times20)$\,$\murm$m$^{2}$ waveguides], on the top of the sinusoidal structure, followed by deposition of 10\,nm Py layer in e-beam evaporator and by the final lift-off. The cross-section of one final structure with 6,000 scans imaged by the scanning transmission electron microscopy (STEM, FEI Helios 660) is shown in Figure \ref{Figure1}. For the spin-wave propagation experiments we additionally patterned 1\,$\murm$m wide, 5/100\,nm thick (Ti/Au) microwave antennas. 
	
	The amplitude of modulation of each structure was measured by Atomic force microscopy (AFM) in order to precisely quantify the effect of the curvature on magnetic properties. The AFM was measured under ambient condition using Bruker Dimension Icon microscope with silicon ScanAsyst-Air cantilevers in ScanAsyst mode. 
	
	\begin{figure*}[ht]
		\centering
		\includegraphics{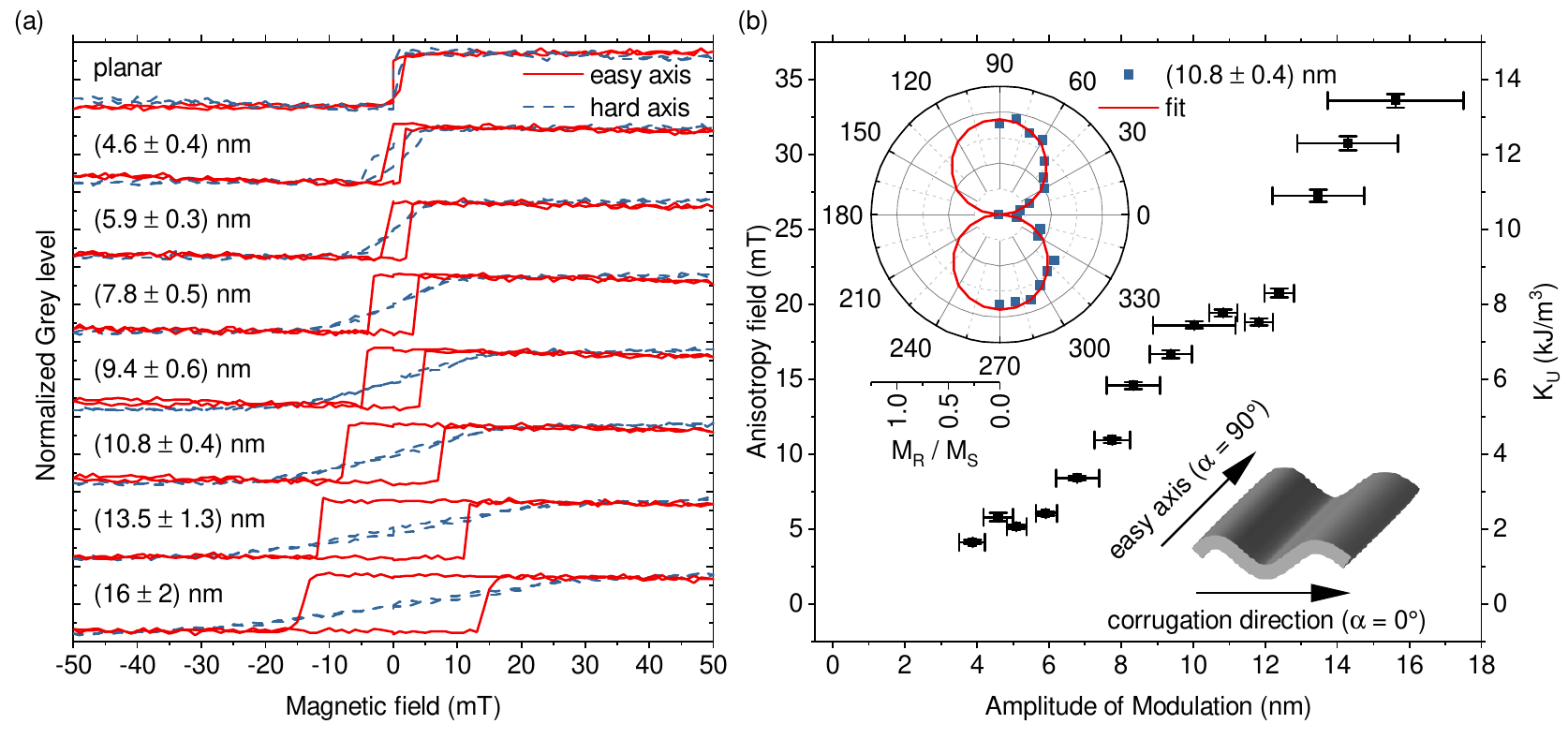}
		\caption{(a) Kerr microscopy measurement of hysteresis loops for the reference planar disc (reference) and for corrugated permalloy discs with increasing amplitude of modulation. The red solid loops were measured in the external magnetic field pointing in the direction perpendicular to the corrugation, the blue dashed loops were measured with the field rotated by $90^\circ$. (b) The dependence of the induced anisotropy field on the amplitude of modulation. The error bars in the amplitude of the modulation axis are obtained from the AFM measurement. The error bars in the anisotropy field axis are extracted from the fit. Inset: Normalized remanence and the fit [using Eq. (1)] of the disc with the amplitude of the modulation of $(10.8\pm0.4)$\,nm depicted as a function of the angle between the corrugation direction and the external magnetic field. }
		\label{Figure2}
	\end{figure*}
	
	The magnetization reversal process and the quantification of the induced magnetic anisotropy were studied in discs with diameter $D=7\,\murm$m by Kerr microscopy (evico Magnetics). The disk shape of the structures studied in this experiment was chosen to avoid any unwanted contribution of the (in-plane) shape anisotropy into the measurement. The in-plane magnetic field was was swept from $-50$ to $50$\,mT. From the acquired images, the average gray level (proportional to the magnetization) of the disks with a varying amplitude of corrugation was extracted as a function of the external magnetic field. The resulting hysteresis loops were normalized to the saturation level. In Figure \ref{Figure2}a, the hysteresis loops for the planar disc and the corrugated discs with seven different amplitudes of modulation are shown. The magnetic field was applied perpendicular and parallel to the corrugation direction. From the shape of the hysteresis loops, we see that the easy axis (dotted curve) lies in the perpendicular direction to the corrugation and the hard axis (solid curve) lies parallel to it. From the shapes of the hysteresis loops, we can see that the uniaxial magnetic anisotropy is becoming stronger with increasing amplitude of modulation. The values of the anisotropy constant $K_\mathrm{U}$ can be calculated using $K_\mathrm{U}=\dfrac{M_\mathrm{S}\mu_0 H_\mathrm{A}}{2}$, where the anisotropy field $\mu_0 H_\mathrm{A}$ can be directly extracted from the hard axis hysteresis loop using piece-wise fit \cite{Colino2016}. The saturation magnetization for the Py films was measured $\mu_0 M_\mathrm{S}=1.04$\,T by vibrating sample magnetometry (VSM, Quantum design Versalab) and confirmed by VNA-FMR experiments \cite{Bilzer2007}. The dependence of the anisotropy field $\mu_0 H_\mathrm{A}$ and the anisotropy constant $K_\mathrm{U}$ on the amplitude of modulation is shown in Figure \ref{Figure2}b. To confirm the uniaxial characteristic of the corrugation-induced anisotropy, we measured hysteresis loops for different in-plane directions of the external magnetic field with respect to the corrugation direction. In the inset of Figure \ref{Figure2}b the remanence $\frac{M_\mathrm{r}}{M_\mathrm{S}}$ as a function of the angle between the corrugation direction $(\alpha=0^\circ)$ and the external magnetic field is depicted for the disc with the corrugation amplitude of 10.8\,nm. This data show clear uniaxial anisotropy and can be fitted using the equation\cite{Kuschel2011}
	\begin{equation}
		\dfrac{M_\mathrm{r}}{M_\mathrm{S}}=\dfrac{M_\mathrm{r}^\mathrm{max}}{M_\mathrm{S}}\cdot\left|\cos(\alpha-\alpha_0)\right|+\dfrac{M_\mathrm{r}^\mathrm{off}}{M_\mathrm{S}}\, .
	\end{equation}
	The fit confirms, that the easy axis is formed perpendicular to the corrugation direction, i.e. at $\alpha=90^\circ$.
	
\begin{figure}[htp]
		\centering
		\includegraphics{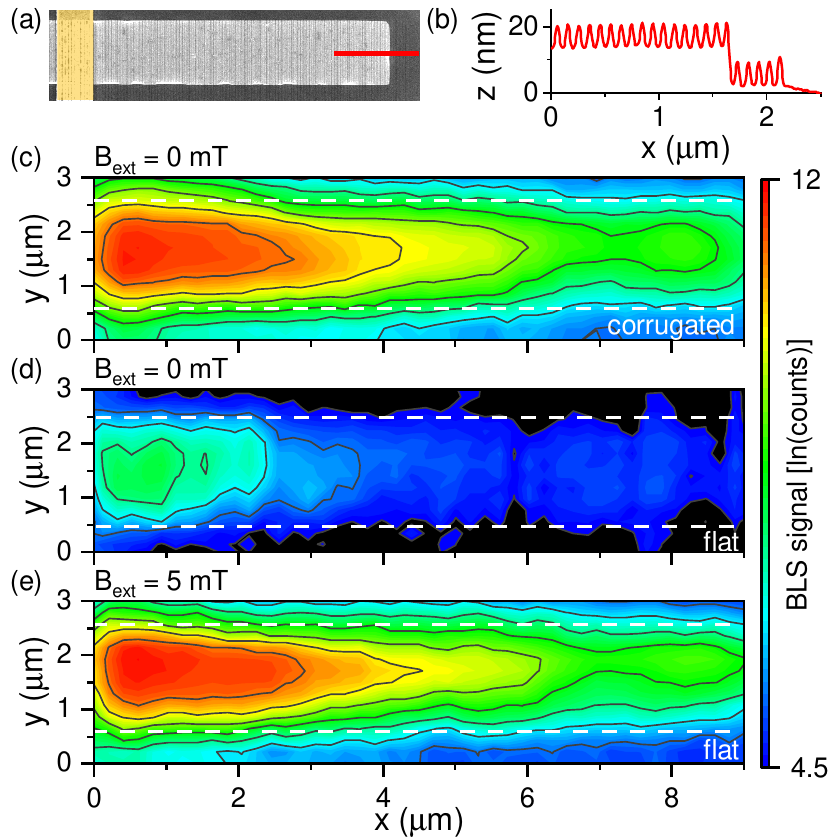}
		\caption{(a) SEM image of the corrugated magnonic waveguide with the excitation antenna (highlighted with ochre color). The red line (scale bar) is $2.5\,\murm$m. (b) Corresponding AFM line profile over the red line of the fabricated structure. The amplitude of modulation (peak-to-peak) for this waveguide is $A=(6.8\pm0.2)$\,nm. The step in the line profile represents the edge of the NiFe magnonic waveguide. (c) BLS intensity map in the logarithmic scale measured on the corrugated waveguide in zero magnetic field with the excitation frequency $f=2.75$\,GHz. Due to the corrugation of the magnonic waveguide, the magnetization vector lies perpendicular to the long axis of the waveguide. Therefore, spin waves propagate in the Damon-Eshbach geometry. The map was recorded over $3\times9\,\murm$m$^2$ by rastering the BLS laser spot with a 200\,nm step. The dashed lines indicate the edges of the waveguide. (d) BLS intensity map in the logarithmic scale measured in zero magnetic field on the planar magnonic waveguide. The excitation frequency was $f=1.8$\,GHz. Due to the shape anisotropy of the waveguide, the magnetization is aligned parallel to the long axis. Hence, the spin waves propagate in the backward volume geometry. e) BLS intensity map in the logarithmic scale measured in the external magnetic field of 5\,mT on the planar magnonic waveguide. The external magnetic field rotated the magnetization perpendicular to the long axis of the waveguide, and the spin-wave propagation geometry changed from backward volume to Damon-Eshbach geometry.}
		\label{Figure3}
	\end{figure}

	To study the spin-wave propagation, we used Brillouin light scattering microscopy (BLS).  This experiment was conducted on the magnonic waveguide with the width $w_\mathrm{w}=2\,\murm$m, length $l_\mathrm{w}=20\,\murm$m and the amplitude of modulation $A=(6.8\pm0.2)$\,nm. The fabrication procedure and the magnetic properties of the waveguide were the same as for the discs used in the previous experiment. To excite spin waves, we placed an antenna with the width $w_\mathrm{a}=1\,\murm$m fabricated on top of the waveguide. The SEM micrograph of the corrugated waveguide with the antenna on top is shown in Figure \ref{Figure3}a. The corresponding AFM height profile is shown in Figure \ref{Figure3}b. The effective field [consisting of corrugation-induced anisotropy contribution (8\,mT) minus the demagnetizing field (3\,mT) coming from the shape of the waveguide] for this waveguide was $\mu_0 H_\mathrm{eff}=5$\,mT. Due to this effective field, the magnetization of the waveguide pointed perpendicular to the corrugation direction (i.e. perpendicular to the long axis of the waveguide) even at zero external magnetic field. To find the optimum spin-wave frequency at zero field we swept the 5\,dBm RF excitation from 1\,GHz to 6\,GHz while measuring the BLS signal at a distance of $5\,\murm$m from the antenna. The maximum BLS signal was measured for the frequency of 2.75\,GHz. At this frequency, we acquired 2D BLS intensity map shown in Figure \ref{Figure3}c. We fitted the BLS intensity integrated in the $y$-direction with the exponential decay $I\cdot\exp(-2x/\delta)$, where $\delta$ is the decay length defined as the distance over which the wave amplitude decreases by a factor of $1/e$. The factor 2 is to consider the BLS intensity being proportional to the square of the spin-wave amplitude \cite{Sebastian2012}. The resulting decay length was $\delta_\mathrm{corrugated}=(5.3\pm0.3)\,\murm$m. 
	
	We then repeated this experiment at zero field for the planar waveguide. Here, due to the shape anisotropy, the magnetization points parallel to the longer axis of the rectangular waveguide. From the frequency sweep, we found that the optimum excitation frequency is 1.8\,GHz. The 2D spin-wave intensity map from this measurement is shown in Figure \ref{Figure3}d. It is apparent that in this case, the decay length is one order of magnitude smaller (approx. \,$0.6\,\murm$m). When we applied an external field to “simulate” the contribution of corrugation induced anisotropy and repeated the BLS measurement at the frequency of 2.75\,GHz, we obtained similar result as for the corrugated waveguide measured at zero field (see Figure \ref{Figure3}e and compare with Figure \ref{Figure3}c). The decay length was evaluated to: $\delta_\mathrm{planar}=(5.7\pm0.5)\,\murm$m.
	
	To further analyze how the amplitude of modulation influences the spin-wave propagation, we prepared an additional sample with higher modulations. We then performed an experiment, where we measured decay lengths for these waveguides. To assure comparable conditions for all waveguides, we fixed the excitation frequency at $f=6.3$\,GHz and varied the external magnetic field in such a way that the spin-wave $k$-vector was kept constant at $k=3.1$\,rad/$\murm$m. These conditions correspond to the effective field of 25\,mT. The $k$-vector was checked before each measurement by performing a phase-resolved BLS line scan\cite{Vogt2009}. This approach ensured that we measured the decay length always at the same point in the dispersion relation and the contribution of the corrugation to the decay length was isolated from the variations in the group velocity. 
	
\begin{figure}[htbp]
		\centering
		\includegraphics{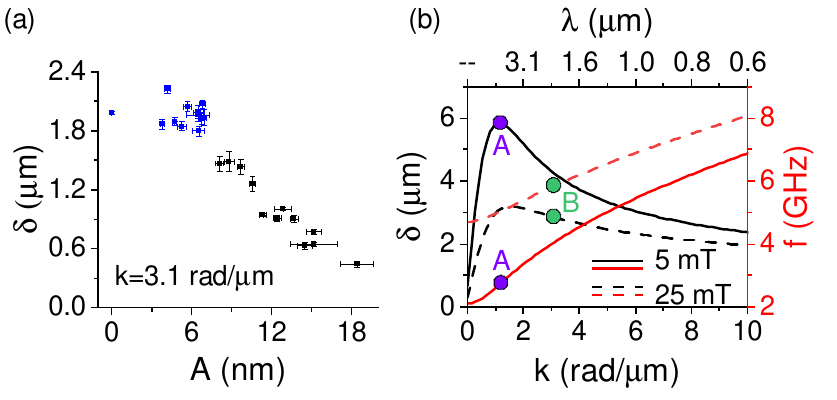}
		\caption{(a) Dependence of the decay length on the amplitude of modulation of magnonic waveguides. The excitation frequency was $f=6.3$\,GHz. To keep the $k$-vector at 3.1\,rad/$\murm$m the external magnetic field was changed for every waveguide with the different amplitude of modulation. (b) Analytical dispersion relations calculated propagation lengths calculated for the first waveguide mode of a transversally magnetized magnonic waveguide. Solid and dashed lines represent results for the external magnetic fields of 5\,mT and 25\,mT, respectively. Point A marks the conditions used in measuring the 2D BLS intensity map in Figure \ref{Figure3}c and \ref{Figure3}e. Point B marks the conditions used for evaluation of the dependence of the propagation length on the amplitude of modulation.}
		\label{Figure4}
	\end{figure}

	Figure \ref{Figure4}a shows the decay length dependence on the amplitude of modulation. From the results, it is apparent that the decay length significantly decreases for the amplitudes of modulation above approx. 8\,nm (Figure \ref{Figure4}a, black points). The modulation of 18\,nm results in the decrease of the propagation length by a factor of three. This decrease can be caused by corrugation-induced inhomogeneities of the internal magnetic field or the inhomogeneities of the magnetic film itself. These inhomogeneities can cause spin wave scattering and can also contribute to the increase of the effective damping; both effects will negatively influence the spin-wave decay length. A positive message is that for low modulations, this effect is negligible (Figure \ref{Figure4}a, blue points). For the planar magnonic waveguide and the waveguides with modulations below 8\,nm, we measured the decay lengths in the range of 1.8 to 2.2\,$\murm$m which corresponds to the propagation lengths of planar Py waveguides reported in other studies \cite{Sebastian2012,Madami2011,Demidov2007}. Note that because these measurements were performed at different $k$-vector, the decay lengths are significantly lower than in the previous experiments. In Figure \ref{Figure4}b we plot analytical calculation \cite{Kalinikos1986} of the dispersion relation (red lines) and propagation length (black lines) for the first waveguide mode of a transversally magnetized planar Py waveguide at two different external magnetic fields. The dispersion was calculated  using the following material parameters: $\gamma⁄(2\pi)=29.3$\,GHz/T, $M_\mathrm{S}=800$\,kA/m, and $\alpha=9\cdot10^{-3}$. The propagation length was calculated as $\delta=v_\mathrm{g}\cdot\tau$, where $v_\mathrm{g}=\frac{\mathrm{d}\omega}{\mathrm{d}k}$ and $\tau=\left(\alpha\omega\frac{\mathrm{d}\omega}{\mathrm{d}\omega_\mathrm{H}}\right)^{-1}$ with $\omega_\mathrm{H}=\mu_0 \gamma H$. The solid and dashed lines in Figure \ref{Figure4}b represent the calculations for the field of 5\,mT and 25\,mT, respectively. Point A corresponds to the conditions in which the 2D BLS intensity map in Figure \ref{Figure3}c was measured. The decay length reaches its maximum (5.8\,$\murm$m) at this point in the dispersion. The experiments where we varied the modulation amplitude were performed at conditions corresponding to point B. The decay length here is approximately 1/2 of the decay length (2.9\,$\murm$m) calculated in point A. These calculations agree reasonably well with the decay lengths obtained from the experiments.
	
	In conclusion, we studied spin-wave propagation in corrugated waveguides grown on modulated surfaces prepared by focused electron beam induced deposition and electron beam lithography. By shaping the magnetic waveguide in the third dimension and by exploiting the curvature induced effects, we managed to locally control the uniaxial magnetic anisotropy. As the spin-wave propagation strongly depends on the angle between the $k$-vector and magnetization, the local control over magnetization direction (without the need for the external magnetic field) is a big step forward towards the realization of complex magnonic circuits. The presented experiments demonstrate the potential of this technique. We have shown, that without the presence of an external magnetic field, spin waves propagate over a distance $10\times$ larger in the corrugated waveguide than in the planar waveguide. Further, we have analyzed the influence of the amplitude of modulation on the spin-wave propagation and concluded that for moderate modulation amplitudes, the spin-wave decay length is not affected. For larger amplitudes, the decay length decreases linearly with increasing modulation. The presented 3D nanofabrication approach is universal and can be used with any commonly used magnetic material. Also, it can be applied to any magnetic system where is a need for local control of the internal magnetic field or magnetization direction. 
	
	This research has been financially supported by Brno University of Technology (grant No. FSI/STI-J-18-5333), by the Technology Agency of the Czech Republic (TN01000008), and by the Czech-Bavarian Academic Agency
	(project no. 8E18B051). CzechNanoLab project LM2018110 funded by MEYS CR is gratefully acknowledged for the financial support of the sample fabrication and measurements at CEITEC Nano Research Infrastructure. I. T. was supported by the Brno Ph.D. talent scholarship.
	
	
	%
	%
	
	%
	
	
	\bibliography{clanek}
	
\end{document}